\documentclass{webofc}
\usepackage[varg]{txfonts}   
\usepackage[autostyle, english = american]{csquotes}
\usepackage{hyperref}
\usepackage{graphicx}
\graphicspath{ {./images/} }

\newcommand\snowmass{\begin{center}\rule[-0.2in]{\hsize}{0.01in}\\\rule{\hsize}{0.01in}\\
\vskip 0.1in Submitted to the  Proceedings of the US Community Study\\ 
on the Future of Particle Physics (Snowmass 2021)\\ 
\rule{\hsize}{0.01in}\\\rule[+0.2in]{\hsize}{0.01in} \end{center}}

\begin{document}

\title{Societal impacts of particle physics projects}

\author{\firstname{Rochelle} \lastname{Zens}\inst{1}
\and
\firstname{Michael} \lastname{Headley}\inst{1}
\and
\firstname{Debra} \lastname{Wolf}\inst{1}
\and
\firstname{Alison} \lastname{Markovitz}\inst{2}
\and
\firstname{Faith} \lastname{Dukes}\inst{3}
\and
\newline
\firstname{Jennifer} \lastname{Tang}\inst{3}
\and
\firstname{Kenneth} \lastname{Bloom}\inst{4}
\and
\firstname{Veronique} \lastname{Boisvert}\inst{5}
}

\institute
{Sanford Underground Research Facility (SURF)
\and
Fermi National Accelerator Laboratory (Fermilab)
\and
Lawrence Berkeley National Laboratory (LBNL)
\and
University of Nebraska-Lincoln, Lincoln, NE 68588
\and
Royal Holloway University London
}

\snowmass

\abstract{
Large particle physics projects funded by the U.S. Government require an evaluation and mitigation of each project’s potential impacts on the local communities. However, beyond meeting governmental requirements, particle physics projects stand to play an essential role in local decision-making, building relationships, and framing discussions about key projects by becoming meaningfully engaged in their local communities. In this white paper for the U.S. Particle Physics Community Planning Exercise (“Snowmass"), we examine several local community engagement efforts made by three facilities:  Lawrence Berkeley National Laboratory (Berkeley Lab), Fermi National Accelerator Laboratory (Fermilab), and the Sanford Underground Research Facility (SURF). Although each facility focuses on a different endeavor in varying types of communities, each study highlights the importance and benefits of employing consistent outreach techniques, promoting diversity, establishing lasting relationships, and creating environments for open and honest communication.
}

\maketitle

\section{Executive Summary}

\label{sec:intro}
\setlength{\footskip}{3.60004pt}

As large employers and leading entities within their communities, particle physics laboratories can benefit from community engagement focused on local impacts. Community engagement plays an essential role in local decision-making, building relationships, and important discussions about the implementation of key projects. Large particle physics projects funded by the U.S. Government require an evaluation and mitigation of each project’s potential impacts on the local communities. Beyond satisfying governmental requirements, lasting and positive change can result when laboratories work alongside their respective communities in a meaningful way, which broadens the positive societal impacts of particle physics research. 

Below are three case studies from Lawrence Berkeley National Laboratory (Berkeley Lab) located in Berkeley, California; Fermi National Accelerator Laboratory (Fermilab) located in Batavia, Illinois; and the Sanford Underground Research Facility (SURF). Each case study presents a community engagement undertaking focused on local impacts distinct to each laboratory. Further, each study highlights the unique circumstances from each of the laboratory’s regions–Berkeley provides an urban perspective to community engagement through its Community Relations and K-12 STEM Education and Outreach programs and through partnering with a local non-profit, Rising Sun, which focuses on workforce development programs for youth. Fermilab presents an example of its suburban approach to community engagement through the work of its Community Advisory Board, while SURF highlights a rural approach to engaging with indigenous groups through education and cultural awareness through the creation of an ethnobotanical garden. 

Although each of the case studies presents a different perspective on community engagement as it interfaces with social impact, several common themes emerge across all three. In all of the studies, employing consistent outreach techniques, promoting diversity, establishing lasting relationships, and creating environments for open and honest communication led to the best outcomes.

\section{Case Study: LBNL Community Impacts}

\label{sec:LBNL}
Lawrence Berkeley National Laboratory (Berkeley Lab) was founded in 1931 by physicist Ernest Orlando Lawrence. Lawrence, who won the Physics Nobel Prize in 1939 for his invention of the cyclotron, believed that the biggest scientific challenges are best addressed by teams and is credited with launching the modern era of multidisciplinary, team science. 

Today, Berkeley Lab ranks among the world’s top research institutions, and conducts basic and applied unclassified research across six key scientific disciplines: biosciences, computing sciences, physical sciences, energy sciences, Earth and environmental sciences, and energy technologies. 

The Lab employs nearly 4,000 staff members, about 1,700 of whom are scientists and engineers who develop sustainable energy and environmental solutions, create useful new materials, advance the frontiers of computing, and probe the mysteries of life, matter, and the universe. More than 500 are undergraduate and graduate students – scientists just beginning their research journey.

A \$1.1 billion science and technology powerhouse, Berkeley Lab, which has been managed by the University of California for the U.S. Department of Energy’s Office of Science since its inception, is still shaping the world in fundamental ways. 

\subsection{Community and Education Outreach at Berkeley Lab}

Berkeley Lab’s Government and Community Relations Office is home to both its Community Relations and K-12 STEM Education and Outreach programs. 

The Community Relations team works to ensure that the Lab is a good neighbor and responsible community institution. Its mission is to serve as a bridge between the Lab and the community by sharing inspiring stories about its ground-breaking research; facilitating opportunities for the Lab to give back to its neighbors; and growing and strengthening partnerships with local governments, nonprofits, and community-based organizations to advance positive social impact in the East Bay, Bay Area, and beyond.

The K-12 STEM Education and Outreach team aims to develop educational programs to teach and support the next generation of scientists, with a goal to connect Berkeley Lab experts with students around the Bay Area, to excite them about the world of STEM, help them explore STEM careers, and give them the experience to pursue STEM in college and beyond.

Both teams work together to advance shared goals around community engagement and have worked to cultivate, grow, and deepen relationships with community-based organizations through a number of methods, which are detailed through the case study below.

\subsection{Rising Sun Center for Opportunity}

The Rising Sun Center for Opportunity (Rising Sun) is a nonprofit workforce development organization based in Oakland, California \cite{bib:risingsun}.  It provides job training and employment programs for youth and adults in the Bay Area and San Joaquin County that support individual and community resilience, combat climate change, and build economic equity. Rising Sun also engages in policy and advocacy work with a focus on integrating workforce development and economic justice into climate solutions, so that a just and sustainable future is possible for all people and the planet. 

Rising Sun’s mission complements the work that Berkeley Lab does in the energy efficiency space and in the STEM education and workforce development space. For that reason, the Lab reached out to the organization in 2011 to explore possible collaborations. As a goodwill gesture to demonstrate the Lab’s interest and commitment to engaging with them, the Lab also made a small philanthropic donation in support of Rising Sun’s Climate Careers youth program, which hires and trains youth ages 15-24 from low-income households to work as energy specialists providing residents in their own local communities with free energy efficiency and water conservation services.

The donation served as a catalyst for engagement, allowing both institutions to come to the table and begin identifying opportunities for the two institutions to work together. Over the next few years, the Lab organized one-off opportunities for Rising Sun’s youth program participants to tour the Lab and learn about potential STEM career pathways from Lab researchers.

Those interactions were well received, and helped build trust between the two institutions. With an interest in collaborating more closely, the Lab furthered its engagement with Rising Sun in 2017 by providing more opportunities for their youth program participants to interact with Lab employees. 

Two specific ways in which the Lab supported Rising Sun’s youth career programs were through career panels and educational design challenges. Prior to the pandemic, the Lab organized in-person career panels with staff in both STEM and STEM support roles. Panelists were chosen based on the diversity of their career pathway, experiences and backgrounds. 

At the conclusion of the youth Climate Careers program, Berkeley Lab staff coordinated with Rising Sun on a visit to the Berkeley Lab site. The full day program included both tours to user facilities like the Advanced Light Source (ALS) and the National Energy Research Scientific Computing Center (NERSC) as well as an energy design challenge. The challenge allowed participants to apply their knowledge gained as energy efficiency experts in their community to a larger design problem. 

\begin{figure}
    \centering
    \includegraphics[scale=0.67]{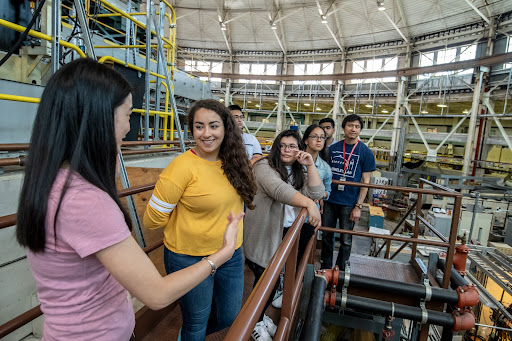}
    \caption{Rising Sun youth program participants tour the Advanced Light Source at Berkeley Lab}
    \label{fig:sec_label}
\end{figure}

In recent years, the challenge has covered the application of LEED building certification requirements in the rehabilitation of a historical building. Working in small groups, challenge participants designed and developed plans that addressed location and transportation, sustainability, water efficiency, energy, and indoor environmental quality. These specific areas directly overlapped with the research of Berkeley Lab’s Energy Technologies Area (ETA) at the Lab. Therefore, many of our volunteer participants have been researchers from ETA, providing guidance to each group throughout the day. The day has been concluded with presentations to selected judges. Judges provide feedback to the groups and ask questions about their thought process, team interactions, and overall experience for the day. The panel has been composed of leadership within the Lab and staff members from our local congressional offices. Certificates from the offices of Congresswoman Barbara Lee and State Senator Nancy Skinner have also been provided to participants upon completion of the day. 

Additionally, the Lab also brought Rising Sun’s President and Chief Executive Officer onto its Community Advisory Group (CAG), which serves as a forum for discussion about the Lab's initiatives and activities, and a venue to identify opportunities for the Lab, City of Berkeley, and other community members to collaborate in support of a vibrant and diverse community. For the last 5 years, Rising Sun has been an integral voice on the Lab’s CAG, providing an important perspective on topics including how to think more equitably about engaging and attracting more people from underrepresented groups into STEM fields.

Through the CAG and other engagements, the Lab and Rising Sun have developed a deep appreciation for, and understanding of, the value of the work each institution does, and have found ways to support and advance each other’s missions. 

As Rising Sun learned more about the Lab and its research priorities, they became a strong third party advocate, weighing in on issues important to the Lab at the state and federal level. For example, it was one of several community-based organizations that encouraged the White House and Congressional leaders to support adequate funding for the U.S. Department of Energy’s Office of Science in the Fiscal Year 2022 President's budget request and the Congressional appropriations process. 

Berkeley Lab and Rising Sun have also joined together with other institutions to pursue state and federal funding opportunities. For example, both are members of a coalition that has been named a finalist in the U.S. Economic Development Administration’s \$1 Billion Build Back Better Regional Challenge.\cite{bib:EDA}  Rising Sun also sits on the Equity Subcommittee for the Lab’s CalFlexHub initiative\cite{bib:calflex}. 

\subsection{Outcomes and Lessons Learned}
Since 2017, more than 600 Rising Sun youth program participants have participated in career panels, more than 60 have participated in the annual design challenge, and last year the Lab hosted two Rising Sun youth program participants for a virtual externship, the first year Rising Sun offered the program. Since 2011, the Lab has donated \$28,500 to Rising Sun to support its programs and mission. 

A few lessons learned have emerged from the Lab’s ten-year relationship with Rising Sun that may be helpful for others as they look to develop relationships with community-based organizations:  
\begin{itemize}
\item Identifying community partners with overlapping missions increases the chances for a more sustainable relationship with shared goals and objectives. 
\item A modest philanthropic donation can serve as one possible entry point for building a relationship with a community-based organization, though to be clear, donations should be one component of a larger engagement strategy. 
\item Building meaningful relationships has to be a two-way street. Oftentimes large institutions like the Lab can feel like it has all the answers, but listening to, and learning from, community-based partners is a critical component of a successful partnership.
\item Sharing time and talent, and exchanging best practices and ideas, provides a strong foundation for trust, which is paramount in any working relationship. 
\end{itemize}

\section{Case Study: Fermilab's Community Advisory Board}

\label{sec:FNAL}
Fermi National Accelerator Laboratory (Fermilab) is a U.S. Department of Energy (DOE) National Laboratory managed by Fermi Research Alliance, LLC. As a premier particle physics and accelerator laboratory, Fermilab seeks to solve the mysteries of our universe at its most fundamental level. Located in Batavia, Illinois, the lab designs and builds world-leading particle accelerators, detectors and computing facilities to dig down to the smallest building blocks of matter and probe the farthest reaches of the universe. Scientists from across the United States and more than 50 countries use Fermilab's scientific facilities, and its education, outreach and tour programs reach approximately 100,000 students and adults annually.

Fermilab was founded in 1967 and is celebrating its fifty-fifth anniversary in 2022. The laboratory has a long history of engaging with its community to share its science and welcome the public on site. The site is 6,800 acres, with much of it as open prairie land; the site is designated as a National Environmental Research Park. Fermilab’s first director, Robert Wilson, brought a herd of bison to the site, which remains today, as a symbol of the history of the Midwestern prairie and the laboratory’s pioneering research at the frontiers of particle physics. Fermilab’s second director, Nobel prize winner Leon Lederman, was committed to physics education. He started the lab’s Saturday Morning Physics program in 1980 and launched education programs that led to the Lederman Science Center, which provides an onsite hub for science education and public engagement. 

Fermilab established its current Community Advisory Board in 2009. The present Board grew out of two former task forces, one charged with developing the laboratory’s overall approach to public participation and one offering specific input on a proposed linear accelerator project. The Board has proven to be an effective means of communicating with the public and obtaining input on activities that may affect the lab’s surrounding communities.

\subsection{Methods (Partners, Approaches, Actions) }

The Fermilab Community Advisory Board (CAB) meets every two months and provides ongoing advice and guidance related to the laboratory’s current activities and future plans. The Board gives feedback on proposed new projects, advises Fermilab on public participation and engagement, and acts as a liaison with local organizations and communities.

Membership on the Board is drawn from the diverse communities of the greater Chicagoland area, including individuals from the surrounding cities as well as from the wider region. Members bring different perspectives – from city representatives, to educators, to scientists, to outdoor enthusiasts, to the science-interested public. As Fermilab is a DOE National Laboratory, the head of the DOE Fermi Site Office participates on the CAB. Senior Leadership from Fermilab also participate in the bi-monthly CAB meetings.

Fermilab’s sustained public engagement over many years has helped to develop and support an engaged and active CAB. Prior to the founding of the CAB, programs like Saturday Morning Physics, the Lederman Science Center, and others brought members of the public to the lab and this connection helped to develop relationships that have led to CAB participation. 

The site’s open outdoor areas, used by the public for bison viewing, biking, running, bird watching, and other activities, also have given local community members an opportunity to come to Fermilab and experience its site. This has also served as a gateway to expose the broader public to Fermilab science. In addition to general outdoor recreation at the lab’s site, the Fermilab Natural Areas organization, formed in 2006, provides an opportunity for volunteers to get involved in restoring, managing and enhancing the natural areas and resources of Fermilab in order to maintain and improve their ecological health and biodiversity. 

Over the years, Fermilab has sought to expand the CAB’s membership and reach different groups. The lab seeks new members every few years through local press releases, which invite interested members of the public to apply. The lab also conducts targeted requests, reaching out to specific groups or civic representatives and asking them to identify candidates to serve on the CAB. Typically, recruitment of new CAB members has worked best where the invited participant has a strong interest in the lab’s work; members who have been appointed without a connection or strong interest in the lab are often less engaged and their participation may wane. Those CAB members with a stronger connection to the lab – through education efforts, personal connections, or a strong interest in the lab’s science or outdoor areas – are typically more engaged. More recently, Fermilab has been developing a Designated Community Representative role for its CAB to help ensure better coordination and communication with local community governments.

Meetings are intended to inform CAB members about key activities of the lab, including areas that may impact the community. Topics include ongoing or planned construction, security and safety reviews, and organizational developments. CAB members are briefed on scientific breakthroughs, major publications, and the progress of the lab’s projects and experiments. CAB members are asked to provide input on the lab’s activities and invited to participate in key community engagement events, such as the annual Family Open House and STEM Career Expo.

\subsection{Outcomes }

The CAB has proven to be an effective way for the community to stay informed and for the lab to receive input from community members on its activities and their potential impacts. CAB members ask probing questions about the status of existing projects and construction activities, they have provided useful feedback on the lab’s planned educational programs, and they have advocated for the community’s views on topics like public access to the site.

One example of the role of the CAB in communicating community impacts is the lab’s tritium management program. Very low levels of tritium were discovered in Fermilab ponds and streams in 2005, well below regulatory limits. Fermilab management interfaced with the predecessor to the CAB (a community group offering input on a proposed linear accelerator project) to communicate about this discovery, answer their questions and seek their input on how to educate the public about tritium. Lab staff conveyed that the low levels pose no threat to human health or the environment. This dialogue helped to guide Fermilab’s response and inform the public. Fermilab continues to provide annual updates to the CAB on the current status of its tritium management; this helps educate new members and provides a regular update to the community on this topic. 

Another instance in which the CAB’s input has been particularly valuable has been when the lab is launching a new project and is engaging with the public regarding potential environmental or community impacts. In 2015, Fermilab and the Department of Energy prepared an environmental assessment for the Long Baseline Neutrino Facility (LBNF) and the Deep Underground Neutrino Experiment (DUNE). A draft Environmental Assessment was released by the Department of Energy and was shared with local residents, public officials, and the media. CAB members were briefed on the planned construction for LBNF at Fermilab’s Batavia site, including construction that would be taking place near the public road and some residential areas. CAB members were shown the planned construction site to help visualize the plans and provided input in connection with the environmental review. There was a public comment period and public meetings were held, in which CAB members participated as well. 

The CAB also has provided the lab with important community ties during COVID.  Since the spring of 2020, the CAB has continued to meet virtually every two months. Lab leadership has kept CAB members informed of access restrictions, safety precautions, and ongoing work on site, and has received feedback from the community including their desire to return to the site. The history of CAB engagement prior to COVID has helped to sustain an active and productive dialogue with the community even as meetings have transitioned to virtual during the pandemic.

CAB members also occasionally participate as volunteers for major Fermilab outreach events. In the summer of 2013, the massive magnet for the Muon g-2 experiment was moved from Brookhaven Laboratory in New York to Fermilab in Illinois. The magnet’s complicated move through waterways and across the country was documented along the way, and a big celebration was held at Fermilab when the magnet arrived at the end of its journey\cite{bib:muon}. CAB members were invited to participate in this event as volunteers; this engagement helped strengthen their connection to the lab and to appreciate the power of its science. As another example, as part of the lab’s fiftieth anniversary celebrations in 2017, Fermilab held a big open house which brought thousands to the lab. CAB members also served as volunteers and ambassadors for the lab.

\subsection{Key Factors for Success / Lessons Learned}

Several factors likely contribute to the successful engagement between Fermilab and its local communities through its Community Advisory Board:
\begin{itemize}
\item CAB meetings are consistently held every two months and the CAB has been in place for over twelve years, building on prior task forces established by the lab to interact with and obtain input from the local communities.
\item Fermilab’s other community engagement activities (student and teacher education programs, public lecture series, internships, open houses, career expos, etc.) help bring in a diverse and committed group to participate as CAB members. Efforts spent on smaller or short-term engagement activities can expand the lab’s profile and grow into these broader and more sustained partnerships. CAB members are typically most engaged when they have a connection to Fermilab beyond participation in the CAB.
\item Representation from local governments has been helpful to facilitate communication and discussion of potential impacts from site activities on local communities. Fermilab has recently been working to strengthen those connections by seeking a Designated Community Representative from each of its surrounding cities to serve on the CAB.
\item It has been important to refresh and expand CAB membership and to identify additional ways to ensure diverse community representation and hear different points of view. A combination of broad-based solicitation, through a local press release, plus targeted requests for specific organizations to identify representatives, has contributed to the diverse membership.
\item In person meetings are preferable for building relationships. The years of in person meetings have helped to sustain active CAB involvement with virtual meetings during COVID.
\item Trust has been built over time through transparent communication and actively seeking input. The lab benefits substantially from two-way communication with its local stakeholders.
\end{itemize}

\section{Case Study: SURF Collaboration with Indigenous Peoples }

The Sanford Underground Research Facility (SURF) is located on the site of the former Homestake Gold Mine in Lead, South Dakota. Until its closure in 2002, Homestake was the largest and deepest gold mine in the Western Hemisphere, producing approximately 41 million ounces of gold in its 126-year lifetime. Nestled in the Black Hills, Homestake’s depth proved to be an ideal location for physics research. Starting in the 1960s, Dr. Ray Davis conducted his solar neutrino experiment at Homestake while the mine was still operational, and later won a shared Nobel Prize in physics in 2002. 

Today, SURF hosts world-leading science experiments in a range of disciplines including, physics, geology, biology, and engineering. SURF provides significant depth and rock stability—a near-perfect environment for experiments that need to escape the constant bombardment of cosmic radiation, which can interfere with the detection of rare physics events.

While SURF’s beginnings stem from the closure of the Homestake Gold Mine, the history of the region dates back to a time well before the discovery of gold in the region. Numerous indigenous groups have considered the Black Hills of South Dakota to be a sacred region since time immemorial. Like the rest of the western United States, indigenous land loss in South Dakota was swift and substantial. From 1851 to 1889, indigenous peoples of the area went from free reign over traditional territories, to being confined to the reservation boundaries we know today\cite{bib:ftlaramie}. Indigenous groups of the region relied heavily on bison, which were a source of food, clothing, tools, and other resources. When Anglo-Americans hunted bison to near extinction during this time period, indigenous populations became heavily reliant on the federal government for rations for their survival. 

This loss of land and traditional economies markedly contributed to cycles of poverty, health conditions, and additional disparities between Native and non-Native populations in South Dakota and beyond that persist today. These disparities and the history that propelled them are two key reasons why the South Dakota Science and Technology Authority (SDSTA) strives to increase its interactions with tribal nations residing within the boundaries of South Dakota. 

SDSTA works alongside tribal groups who reside in South Dakota in several capacities. Some of the efforts have been mandated by federal requirements, including work outlined in the National Environmental Policy Act and the National Historic Preservation Act to consult tribes about the relevant work at SURF. Other efforts have been self-initiated to further partnerships with the Tribes, two of which are outlined below. 

\subsection{Education and Outreach at SURF }

The Education and Outreach team at SURF works with K-12 schools and post-secondary learning institutions across South Dakota. Their efforts include leading field trips to SURF, presentations at schools, development and deployment of no-cost STEM curriculum units to classrooms across the state, and teacher professional development and support. As part of the team’s outreach efforts, the Education and Outreach team participates in the annual South Dakota Indian Education Summit. The summit has served as an event for educators of Native children to convene and share best practices since 2004\cite{bib:bonnett}. The Education and Outreach team first met with teachers from Isna Wica Owayawa or Loneman Day School at the 2019 Summit. 

\subsection{Mni Wiconi - Water is Life and Working with a Tribal School }

Loneman School is located in Oglala, South Dakota. Oglala is located within the Pine Ridge Indian Reservation, land which was designated for the use and occupation of the Oglala Lakota people. Isna Wica Owayawa is Oglala’s only school. The median household income of Oglala Lakota County was \$17,300 in 2019, with a poverty rate of 55.8 percent\cite{bib:census}. Isna Wica Owayawa serves approximately 250 students in kindergarten through eighth grades.

At the South Dakota Indian Education Summit, SURF’s Education and Outreach team provided attendees with information about summer professional development and began the work of scheduling a full day of programming for all of the teachers at Isna Wica Owayawa. Although scheduling conflicts and other delays impacted the timing of the training, three teachers from the school attended a professional development training in the summer of 2020, which was held virtually due to the COVID-19 pandemic. Following the professional development, one of the teachers asked the team about developing supplemental material for an after-school program, held monthly on Saturdays for both students and parents.
The Education and Outreach team began their work by asking students what they were interested in learning more about, with the older students expressing interest in learning more about the environment\cite{bib:lone}. The Education and Outreach team then decided to adapt an existing presentation, ''Water: Where Does It Go?''to be more culturally and geographically relevant to the school and its students. 

To begin, the program was renamed Mni Wiconi, which translates from Lakota to “Water is Life.” Mni Wiconi is also associated with the name of the largest tribal-rural water project in the United States, which provides water to a 12,500 mile area, encompassing both rural and tribal areas across the state\cite{bib:mni}. The phrase was also used frequently during the protests at Standing Rock during construction of the Dakota Access Pipeline, as Lakota beliefs state that “the Mni Oyate, the Water Nation. She is Alive. Nothing owns her. Thus, the popular Lakotayapi assertion “Mni Wiconi”—water is life or, more accurately, water is alive.”\cite{bib:standing}

With the cultural importance of water in mind, the lessons also placed emphasis on a nearby water source, the White River, which many local residents believed to be highly polluted due to its cloudy nature. Staff at SURF tested the water, and found that the water’s cloudy appearance revolved around sediment. The results were relayed with both students and adults of the school. 

To highlight science at SURF, the Education and Outreach team presented videos of the underground area, and asked students to identify potential issues that could occur with the surface water as a result of the research. Staff then discussed the importance of SURF’s wastewater treatment facility, and the role that it plays in ensuring the local water supply is not negatively impacted by work at the lab. 
Finally, to provide broader connections, the Education and Outreach team explored the impacts of fertilizers and excess nutrient run-off in the northern United States, and their contributions to dead zones in the Gulf of Mexico. 

Adaptation and incorporation of Lakota cultural practices into the Mni Wiconi lessons is far from complete. As the Education and Outreach team moves forward, they plan to work alongside a former SURF intern who was also an alumna of Isna Wica Owayawa to further the cultural work with the curriculum unit. Possibilities include incorporating more cultural practices surrounding water and providing an in-person tour of SURF’s wastewater treatment facility. 

\subsection{Cangleska Wakan - The Sacred Circle Garden }

\begin{figure}
    \centering
    \includegraphics[scale=0.7]{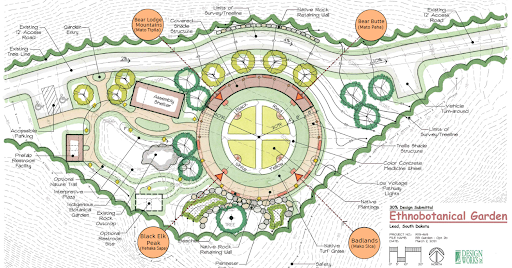}
    \caption{Planned Sacred Circle Garden design}
    \label{fig:my_label}
\end{figure}

As mentioned previously, the Black Hills region holds a sacred significance to many indigenous groups from the region. Cangleska Wakan, or the Sacred Circle Garden, will be located on a hilltop meadow on SURF’s property that will be developed into an ethnobotanical garden. The garden will serve as a space to create awareness of indigenous cultures of the region. Cangleska Wakan is currently slated to be completed in the fall of 2023. The garden is being funded by private donations through the SURF Foundation. The concept, design, and other logistics of the garden are supported by teams across SURF including the SURF Foundation, Education and Outreach, Inclusion, Diversity and Access (IDEA), and Communications. 

The teams have been presented with several challenges during the design and fundraising phases of Cangleska Wakan. One of the first challenges encountered by the teams was placement of the garden. Designers of the garden had little experience with Native cultures, and initially placed the sacred circle portion of the garden in a place that made sense in terms of design, but did not place the circle predominantly in view, which was a major consideration from a Native design perspective. SDSTA staff consulted with local indigenous elders, Richard Moves Camp and Jace DeCory, a professor emeritus at nearby Black Hills State University. Several iterations of the design were drafted until all parties were pleased with the outcome. 

Another challenge was presented in the development of a fundraising video for the Sacred Circle Garden. After consulting members of the Cultural Advisory Committee, a group that advises the SDSTA on cultural efforts at SURF, it was determined that the individuals featured in the video were mispronouncing Cangleska Wakan. After working with members of the Cultural Advisory Committee, the team redid the audio portions of the video with a pronunciation that was more accurate. 

Finally, cultural sensitivity issues related to the garden were raised by a community member regarding images placed on the Sacred Circle Garden’s webpage. One of the photographs included prayer ties near the summit of Black Elk Peak, an area considered to be sacred by the Lakota and which is host to the Welcoming of the Thunders ceremony each spring. After consulting with the community member, the image was replaced with another which was more culturally appropriate. 

\subsection{Factors for Success}

Building a meaningful relationship with the Isna Wica Owayawa community was key to SURF’s successes with the school. Several approaches to successful relationship building were taken when working with Isna Wica Owayawa, including taking the time to learn how the community functions. Team members learned early on that approaches to scheduling meetings and establishing programming varied from SDSTA’s normal operations. Additionally, SDSTA members identified informal leaders in the school community, by asking interested teachers what they needed or wanted, rather than offering lessons that had already been developed. By working with teachers and students at Isna Wica Owayawa to identify areas of interest in specific curricula, SDSTA team members were better able to address the community’s needs on their own terms. Finally, SURF has avoided a major pitfall of relationship building by avoiding a transactional relationship.\cite{bib:UWE} The relationship with Isna Wica Owayawa has grown from an offering of professional development for teachers, to a tailored curriculum unit for their students. Future work with the Mni Wiconi lessons are in development, and opportunities for fully-funded field trips to SURF are being planned. 

Collaboration with culture experts and working with diverse teams helped ensure that both Cangleska Wakan and the Mni Wiconi curriculum were developed with cultural knowledge at the forefront of consideration, not as an afterthought. With Mni Wiconi, feedback from the students at the school regarding what they wanted to learn about contributed to a greater overall engagement in learning both about water systems and the relevant science at SURF. With the ethnobotanical garden, diverse voices worked together to find solutions to cultural concerns in the design of the garden. While some of these discussions proved to be difficult and were at times tense, the overall result was elevated as a result of the collaboration. Recent research supports this outcome. While homogenous teams experience greater levels of comfort, those with differing backgrounds and identities produce better outcomes since they must navigate new and uncharted perspectives to arrive at a shared goal.\cite{bib:HBR}

In both of these examples, SDSTA team members needed to adapt to unfamiliar situations. Having patience, humility, and a willingness to learn and work with diverse teams were essential to the successes experienced with these projects.

\section{Recommendations}

\label{sec:recs}

We offer the following recommendations to laboratories and other facilities who are looking to begin or expand their community engagement efforts. 
\begin{itemize}
    \item Laboratories should engage with their local communities in order to create awareness about their work and build lasting, positive relationships. Community engagement plays an essential role in local decision-making, building relationships, and important discussions about the implementation of key projects. Large particle physics projects funded by the U.S. Government require an evaluation and mitigation of each project’s potential impacts on the local communities. In addition to satisfying governmental requirements, working alongside their local communities can foster lasting change that broadens the positive societal impacts of particle physics research.
    \item Laboratories should have consistent outreach and engagement efforts that provide regular opportunities for feedback to help establish trust. Through its Community Advisory Board, Fermilab offered regularly scheduled meetings to gain feedback from local communities. In addition, SURF ensured its communication with stakeholders at Isna Wica Owayawa was consistent and persistent in order to overcome scheduling and other barriers. 
    \item Laboratories should promote diversity of membership and collaborative efforts in their outreach initiatives to bring a variety of perspectives to the table and create a better end project. SURF’s work with tribal elders and other leaders in its local community helps ensure perspectives of indigenous populations in the region are represented and reflected in the work of the Sacred Circle Garden. Meanwhile, Fermilab regularly refreshes and expands its CAB membership to ensure it remains representative of the diversity of its suburban area.
    \item Laboratories should avoid transactional relationships when developing relationships with stakeholders, and instead focus on approaches that provide value to each entity. Laboratories will be best served by making an extended commitment to working with collaborators over an extended period of time, rather than one-time interactions. Opportunities to receive feedback and consider changes can have lasting impacts on the collaborative efforts. SURF has continued to see improvement in program outcomes with Isna Wica Owayawa using this approach. Berkeley Lab has seen success by utilizing small investments in staff time, small-scale donations, and other resources as a launch pad for lasting collaborations with organizations with shared goals and values. 
    \item Laboratories should utilize methods that promote honest, two-way communication when engaging in collaborative efforts with stakeholders. All three case studies exemplify the benefits of open communication. The CAB at Fermilab creates a space where local community members and the lab are able to air concerns and discuss solutions. Berkeley Lab ensures that its community engagement interactions provide a space for members of the community and partners to voice their opinions, while Berkeley Lab listens and reflects on the opinions shared. Finally, SURF seeks indigenous perspectives although in some instances, the resulting dialogue can result in uncomfortable conversations. However, by promoting difficult conversations in a safe environment, SURF was able to promote a design for its ethnobotanical garden that was approved by all involved.
\end{itemize}
\newpage

\begin{thebibliography}{}
\bibitem{bib:lone} The after-school program was offered to students in kindergarten through eighth grade. Programming was divided into two levels: Kindergarten through fourth grade, and fifth grade through eighth grade.
\end{thebibliography}
%
%

\end{document}